\documentclass[preprint]{elsarticle}

\pdfoutput=1

\usepackage{lineno,hyperref}
\usepackage{listings}
\RequirePackage{amsmath,amssymb,ifthen,url,graphicx,color,array,theorem}

\journal{}

\biboptions{authoryear}

\begin{document}

\begin{frontmatter}

\title{Multi-step Uniformization with Steady-State Detection in Nonstationary M/M/s Queuing Systems}

\author[rvt]{Burak Maciej}
\ead{ maciej.burak@zut.edu.pl }
\address[rvt]{Applied Informatics, West Pomeranian University of Technology,ul. Sikorskiego 37, Szczecin, Poland}




\begin{abstract}
A new approach to the steady state detection in the uniformization method of solving continuous time Markov chains is introduced. The method is particularly useful in solving inhomogenous CTMC's in multiple steps, where the desired error bound of the whole solution can be distributed not proportionally to the lengths of the respective intervals, but rather in a way, that maximizes the chances of detecting a steady state. Additionally, the convergence properties of the underlying DTMC are used to further enhance the computational savings due to the steady state detection. The method is applied to the problem of modeling a Call Center using inhomogenous CTMC model of a M(t)/M(t)/s(t) queuing systems.
\end{abstract}

\begin{keyword}
\texttt{  Markov processes,  OR in service industries,  transient solutions, uniformization, numerical methods, nonstationary, algorithms, Contact Centre} 
\end{keyword}

\end{frontmatter}


\section{Introduction}
One of the most important characteristics of the telephone Call Centers is their varying number of service requests (calls) in time. As the cost of labor is the most significant one in such service systems, the problem of adequate scheduling of its employees has a long history in the area of operational research. The methods proposed to analyze such time-varying behavior in order to find optimal working schedules for given demand forecasts are based mostly on approximations, by adopting stationary queuing models. Examples of such well established methods can be found e.g in \cite{Green_2007coping} or \cite{Aksin_2007}.

However, stationary solutions for generating such schedules are, in many situations, not adequate. \cite{Ingolfsson_2010} demonstrated that such approximations can be either entirely unreliable or deliver results significantly different than methods based on inherently transient models. Despite this, their widespread use is commonly justified by simple implementation and low computational cost.
Another inspiration for this research was the problem of real-time schedule adjustments in situations when we have to deal with much higher than planned volume of service requests, technical problems resulting in higher service times or staff absence. In order to maintain both service level and efficiency, such decisions can involve moving of the break times, extending shift times or temporarily involving  other employees (quality assurance, supervisors) in answering the calls. Although usually generating higher cost than the normal schedule (paying overtime or involving higher paid personnel), requiring very short decision time and being rather common than exceptional, they are, surprisingly, not particularly well supported by current, otherwise quite sophisticated, Call Center management systems. As they are transient by their very nature and often deal with an overloaded system, their modeling using stationary queuing models as proposed, for example, in \cite{Mehrotra_2009} would deliver possibly unreliable results as well.

The main objective of this work is to model such non-stationary systems in a reliable and precise way with computational efficiency enabling its use for schedule planning and in real-time applications.

The paper is structured as follows. In the next section the model and the basic notation are introduced. Section 3 introduces continuous time Markov chains. Section 4 reviews the original uniformization algorithm with steady-state detection and section 5 introduces its modifications for modeling inhomogenous M/M/s queuing systems with particular emphasis on the proposed new steady state detection and error control methods. The paper ends with a summary of experimental results, implementation details, conclusions and proposals for future research.

\section{Model}
To demonstrate the implementation we use the following simplified model of a Call Centre: the analyzed period is finite (e.g. one working day) with the system starting empty; the size $n(t)$ of the system which represents the number of possible states is finite, equal to $s(t)$ = number of servers plus $q(t)$ = capacity of the queue, with corresponding discrete state space $\varphi = \{0,..,n\}$,$|\varphi(t)|=1+s(t)+q(t)$, representing number of service requests (served/waiting calls) in the system.
Customers arrive according to an inhomogenous Poisson process with rate $\lambda(t)$, the service rate $\mu(t)$ is exponential. The load $\rho = \lambda(t)/s(t) \mu(t)$ can be bigger than 1.
There is no abandonment, therefore, the capacity of the queue has to be big enough to be considered practically infinite, which is insofar realistic, as the cost of setting practically unlimited queue space in the telecommunications equipment is negligible nowadays.
The system size must, in consequence, ensure that the probability of being in the state $n$ (abandoning service requests) is insignificant compared to the required computational precision of the whole model, effectively approximating an M/M/s system.

\section{Inhomogenous CTMC}
The description and notation is based on \cite{Haverkort_2001} and  \cite{Arns_2010}.
A CTMC is described by \emph{infinitesimal generator matrix} $Q:n+1 \times n+1, Q=(q_{i,j})$
and the \emph{initial state probability vector} $p(0)$,
where the value $q_{i,j}(i \neq j)$ is the rate at which the state $i$
changes to state $j$ and
$q_{i,i}=-\sum_{j \neq i} q_{i,j}$
represents the  rate for the event of staying in the same state.
The transient distribution at time t $p(t)$ can be calculated using Kolmogorov`s forward equations:
\begin{equation}\label{qh} p`(t)=p(t)Q \end{equation}
Where the vector $p(t)=[p_{0}(t),...,p_{n}(t)]$ gives probabilities of the system being in any of the states at time $t$.

In the inhomogenous case, the generator matrix changes with time $t$ and is denoted as $Q(t)$. Consequently the transient distribution $p(t)$ in that case will take the form:
\begin{equation}\label{qih} p`(t)=p(t)Q(t) \end{equation}
Both represent linear systems of ordinary differential equations, which can be solved using either direct methods for (\ref{qh}) or using numerical approximations for both to solve them in steps. An overview of available numerical solvers can be found in \cite{Oelschl_gel_1990} or \cite{stewart1994introduction}, many of them are available e.g. in the GNU scientific library. Particularly interesting in this context are the embedded Runge-Kutta methods, producing two approximations of different order (e.g. 4+5 in popular Runge -Kutta Fehlberg or 8+9 in Runge-Kutta Prince-Dormand). The comparison of the two solutions allows direct estimation of the error of the approximation  for the given size of the computational step and can be alternatively used for adaptive step size control for the required error bound.

\section{Uniformization}
Uniformization or Randomization, known since the publication of Jensen in 1953 and, therefore, often referenced as Jensen method, is the method of choice for computing transient behavior of CTMCs. Many authors compared its performance in different applications with the conclusion that it usually outperforms known differential equation solvers (e.g. \cite{Grassmann_1978}, \cite{Reibman_1988}, \cite{Arns_2010}). 
To use uniformization we first define the matrix
\begin{equation}
\label{pmx}
P=I+\frac{Q}{\alpha}
\end{equation} which for $\alpha\geq max_{i}(|q_{i,i}|)$ is a stochastic matrix. The value of $\alpha$ is called uniformization rate. Further, let
\begin{equation}
\beta(\alpha{}t,k)=e^{-\alpha{} t} \frac{(\alpha{}t)^k}{k!}
\end{equation}
be the probability of a Poisson process with rate $\alpha$ to generate $k$ events in the interval $[0,t)$. One now finds for $p(t)$
\begin{equation}
\label{jensen}
p(t)=p(0)\sum_{k=0}^{\infty}\beta(\alpha{}t,k)(P)^k
\end{equation}
The formula (\ref{jensen}) can be interpreted as a discrete time Markov process (DTMC) embedded in a Poisson process generating events at rate $\alpha$.

The implemented uniformization algorithm is based on \cite{Reibman_1988}
and computes transient state probabilities for a CTMC  with the following modification of (\ref{jensen}) :
\begin{equation}
\label{reibmann88}
p(t)=\sum_{i=0}^{\infty}\Pi(i)e^{-\alpha t} \frac{(\alpha t)^{i}}{i!}
\end{equation}
where  $\alpha$ is uniformization rate, as described in (\ref{pmx}), and $\Pi(i)$ is the state probability vector of the underlying DTMC after each step $i$ computed iteratively by:
\begin{equation}
\label{reibmann88pi}
\Pi(0)=p(0),\  \Pi(i)=\Pi(i-1)P
\end{equation}
To compute $p(i)$, within prespecified error tolerance, in finite time, the computation stops, when the remaining value of cdf of Poisson distribution is less than the error bound $\epsilon$:
\begin{equation}
\label{reibmann88k}
1-\sum_{i=0}^{k}e^{-\alpha t} \frac{(\alpha t)^{i}}{i!} \leq \epsilon
\end{equation}
with $k$ being the \emph{right truncation point}.
As $\alpha t$ increases, the corresponding probabilities of small number of $i$ Poisson events occurring become less significant. This allows us to start the summation from the $l$`th iteration called \emph{left truncation point} with the equation \ref{reibmann88} reduced to:
\begin{equation}
\label{reibmann88truncated}
p(t)=\sum_{i=l}^{k}\Pi(i)e^{-\alpha t} \frac{(\alpha t)^{i}}{i!}
\end{equation}
 \citep{Reibman_1988} suggests that the values of $l$ and $k$ be derived by:
\begin{equation}
\label{reibmann88lk}
\sum_{i=0}^{l-1}e^{-\alpha t} \frac{(\alpha t)^{i}}{i!} \leq \frac{\epsilon}{2},\ 1-\sum_{i=0}^{k}e^{-\alpha t} \frac{(\alpha t)^{i}}{i!} \leq \frac{\epsilon}{2}
\end{equation}

The main computational effort of the algorithm lies in consecutive $k$  matrix vector multiplications (MVM), necessary for calculation of epochs of DTMC in (\ref{reibmann88pi}) and is of $O(\eta k)$ where $\eta$ is the number of nonzero elements of (sparse) $P$. As our M/M/s/n model is a birth-death process and only transitions between neighboring states are possible, the resulting transition matrix will be tridiagonal ($\eta=3n$) of size $n=s+q+1$ with $max_{i}(|q_{i,i}|)=\lambda + s\mu$. 

As in the case of a Call Center, the optimal setting of the system in terms of the trade-off between service quality and efficiency is achieved for approximately steady load $\rho=\frac{\lambda}{s\mu}$, the number of (scheduled) servers will then grow approximately linearly with $\lambda$. 
Further, if we assume the proportion of servers to the queue capacity to be constant and being some fraction of system size $s=\nu n$, the resulting value of uniformization rate will be $\alpha\geqq\rho n \nu (1+\mu)$. The computational complexity dependent of the size of the analyzed system would be then roughly of:

\begin{equation}
\label{complexity}
O( n^2 \Theta t),\ \  \Theta\thickapprox 3\rho \nu(1+\mu)
\end{equation}

For large $\alpha t$, as the distribution converges to normal, which is symmetric with the mean $\alpha t$, both left and right truncation points $l$ and $k$ in (\ref{reibmann88lk}) will tend to be symmetric to the mean. The number $\frac{l+k}{2}$ is consequently of $O(\alpha t)$ and the number of additional $\frac{k-l}{2}$ MVMs for the given error tolerance  of $O\sqrt{\alpha t}$ and proportional to inverse cdf for that given $\epsilon$.
Therefore, although we could solve the $p(t)$ with any accuracy $\epsilon > 0$, choosing a higher, acceptable for respective practical application, value would mean some computational advantage.

The savings due to (tighter) left truncation are, however, rather insignificant, unless, as proposed in \cite{Reibman_1988} the computation of the first significant DTMC is performed in an improved way. 

An example for this could be precomputing of a $(P)^k$ in (\ref{jensen}) by successive squaring of $P$ which were practical for moderate system sizes or systems with low sparsity as it causes a fill-in of subsequential transition matrices.

Another method to compute the DTMC vectors in a more efficient way, presented first in \cite{muppala1992numerical}, is based on recognizing the steady-state of the underlying DTMC. If convergence of the probability vector in (\ref{reibmann88pi}) is guaranteed then we can stop the MVM after arriving at the steady-state. We can state their proposal as follows:
let us assume that DTMC has the steady state solution $ \Pi (\infty )$,
and that after the $S$ iteration of (\ref{reibmann88pi}), $\Vert\Pi(S) - \Pi(\infty)\Vert_v < \delta(S)$, where $\Vert.\Vert_v$ is an arbitrary vector norm. Then (\ref{reibmann88truncated}) changes to:
\begin{equation}
\label{muppala92}
\hat{p}(t)=
\begin{cases}
\Pi(S) & \text{if } S\leq l,\\
 \\
\displaystyle{\sum_{i=l}^{S}\Pi(i)e^{-\alpha t} \frac{(\alpha t)^{i}}{i!} + \Pi(S)(1 - \sum_{i=0}^{S}e^{-\alpha t} \frac{(\alpha t)^{i}}{i!})}& \text{if } l<S \leq k,\\
 \\
\text{same as }p(t) \text{ in }(\ref{reibmann88truncated}) & \text{if } S>k
\end{cases}
\end{equation}
with $\hat{p}(t)$ used instead $p(t)$ denoting transient state probability vector computed using approximate steady state DTMC vector $\Pi(S)$.
According to \cite{Malhotra_1994} for a predefined error bound $\epsilon$ (as in \eqref{reibmann88k},\eqref{reibmann88lk}) the following inequality holds:

\begin{equation}
\label{malhotra94error}
\ \Vert p(t) - \hat{p}(t) \Vert < \frac{\epsilon}{2} + 2\delta (S)
\end{equation}

The computing of consecutive epochs of the DTMC is equivalent to the power method of finding stationary probability vector of a finite Markov chain. According to \cite{stewart2009probability}, if the stochastic matrix $P$ is aperiodic, convergence of the power method is guaranteed and the number of iterations $k$ needed to satisfy a tolerance criterion $\xi$ may be obtained approximately from the relationship 

\begin{equation}
\label{stewart2009}
\ \rho ^k = \xi  \text{, i.e.,  }  k=\frac{log \xi}{log \rho}
\end{equation}

where $\rho$ is the magnitude of subdominant eigenvalue $\lambda _2$ of matrix $P$ 
\[ 1=\Vert \lambda _1 \Vert > \Vert \lambda _2 \Vert \geq \Vert \lambda _3 \Vert ... \geq \Vert \lambda _N \Vert \]
reducing, consequently, the computational complexity to $O(\eta \  log \xi / log \vert \lambda_2 \vert )$. The detailed analysis of convergence of the power method may by found in \cite{O_Leary_1979} or in standard books on numerical analysis.

Since in most cases the size of the subdominant eigenvalue is not known in advance, the usual method of testing for convergence is to examine some norm of the difference of successive iterates:
\[ \Vert \Pi_i(k) - \Pi_i(k-m) \Vert < \xi \]
 \cite{stewart2009probability} recommends using the relative convergence test of iterates spaced apart by $m$ being function of the rate of convergence:

\[ max_i\left( \frac{\vert \Pi_i(k) - \Pi_i(k-m) \vert}{\vert \Pi_i(k) \vert} \right) < \xi \] 

and suggests envisaging further different convergence tests in order to accept the approximation as being sufficiently accurate.

\subsection{Uniformization for ICTMCs}
The infinitesimal generator matrix $Q(t)$ of an inhomogenous continuous-time Markov chain (ICTMC) is time dependent and the process is described by modified Kolmogorov`s forward equations \eqref{qih}.

As it is still of the type $y'=f(t,y)$, with $y(t)=p(t)$ and $f(t,y) = Q(t)y$, it could be solved by using iterative ODE solvers (e.g. Euler or the already mentioned Runge-Kutta methods) as demonstrated, for example, in \cite{Arns_2010}.

When the changes in generator matrix Q occur in a discrete way at finite points of time and all the rates are constant during the intervals between them, we could replace the analyzed ICTMC with a sequence of homogeneous systems computing the state probability vectors for consecutive time periods recursively using uniformization (e.g. as in \cite{Gross_1984}).

In case of a Call Center, time dependent changes in $Q$ can occur either discretely due to the changing number of servers (arrivals or departures of agents due to the schedule, planned and unplanned breaks, after call work or technical errors like failures of working places etc) or due to changes in the arrival rate.

Since the forecast and current traffic data in Call Center Management applications are already aggregated with their average values by some arbitrary period (e.g. 5, 15 or 30min), we will further assume, similarly to \cite{Ingolfsson_2010}, Q(t) being accordingly piecewise constant and refer to such consecutive time periods with the coresponding HCTMCs as steps.

For the implemented model a pre-emptive discipline is assumed, where the service request rejoins the queue when all servers are busy and a server number decrease happens. This allows for reuse of the calculated probability vector for the next period without modification, which changes only the  interpretation of probabilities of a state from being served to waiting (or the opposite for the server increase). The detailed description of an alternative approach (the exhaustive discipline) can be found e.g. in \cite{Ingolfsson_2007} where servers finish serving requests in progress.

Another approach adopting uniformization for time-inhomogenous CTMCs introduced by \cite{van_Dijk_1992} with subsequent improvements by \cite{van1998numerical}, \cite{Arns_2010} and \cite{andreychenko2010fly} could be used if continuous arrival rates were available, reducing the error of the approximation with the average rates.
 
An additional aspect in case of  a Call Center is that for long time steps even a very exact calculation of the final state probability vector is not sufficient if some performance indicators, like service level or any other distribution of waiting times, are to be estimated as in \cite{Green_2007} or \cite{Ingolfsson_2010}. For this purpose additional intermediary results are needed and that, as in the most transient cases they cannot be simply interpolated with required precision, implies the need for additional splitting of long calculation steps accordingly.

\section{Multi-Step Uniformization with steady-state detection}
The main purpose of modeling a Call Center is the prediction of service level as a function of time. The service level is defined in practical application as percentage of customers, waiting for being serviced longer than $d$ time units (e.g. SL of 90/10 means that no more than 10\% of service requests would have to wait longer than 10s). The exact expressions how to calculate predicted SL as a function of consecutive state probability vectors and number of servers can be found in \cite{Green_2007} or \cite{Ingolfsson_2010}. In order to estimate correct values for SL, the error of calculated state probabilities should stay continuously within the predefined range for the whole modeled period.

One of the biggest advantages of the uniformization is its strict error bounding for one step independently of its length. However, to model an ICTMC according to the above requirements we have to control the error of computation for a number of steps with possibly different lengths.
It is not difficult to show (e.g. \cite{de_Souza_e_Silva_2000}) that the total error for a number of uniformization steps is the sum of truncation errors (error bounds) for each step, consequently the total error bound has to be distributed to sub-intervals.

Assume for a time period $T$ with a known initial distribution $p(0)$ that for any $p(\tau)$, $\tau=(0,T]$ the value of each its state has to be computed with an error less than $\varepsilon_T$. Let us further assume $\varepsilon_t < \varepsilon_T$ being the error after computing some $p(t), t<T$. Then:
\begin{equation}
\label{silvaepsilon}
 \varepsilon_t + \sum_i {\epsilon_{\Delta_i}}  \leq \varepsilon_T ,\  \sum_i {\Delta_i} = T-t
\end{equation}

According to \cite{Arns_2010} for $\varepsilon_R=\varepsilon_T-\varepsilon_t$ being the remaining error, in a step of length $\Delta \leq (T-t)$ starting with $p(t)$ the error should be
\begin{equation}
\label{arnsepsilon}
\epsilon_\Delta \leq \varepsilon_R \frac{\Delta}{T-t}
\end{equation}
to not exceed the error $\varepsilon_T$. This implies distribution of the error bound proportional to the length of the respective single interval. Although it is very intuitive, one could also consider, according to the already mentioned computational complexity of higher right truncation values, which is asymptotically of $O\sqrt{\alpha t}$, to set rather higher error bounds for the steps with smaller $\alpha t$ (shorter size or lower activity) or, as  explained further in the text, for higher steady-state detection thresholds.  For example, tightening the error bound from $10^{-7}$ to $10^{-13}$ requires for $\alpha t = 255$  about 20\%  of additional MVM, whereas for an $\alpha t = 4095$ the increase is only 6\%.

\

As the DTMC representing an M/M/s/n is both irreducible and aperiodic, its limiting (steady-state) distribution is unique and independent of the initial distribution.
The convergence properties required for existence of a steady-state solution of queuing systems can be found e.g. in \cite{stewart2009probability}. Due to the fact that the model does not allow abandonment and approximates an infinite capacity system with a M/M/s/n system of appropriate capacity, the existence of steady-state (in the meaning of either $n$ not growing infinitely or for a $n$ fixed -  $\forall{t}: p_n(t)$ not bigger than some arbitrary value limiting adequacy of the approximation) requires  $\lambda<\mu s$ as for an M/M/s system (such system will further be referred to as converging). Its precise stationary distribution $\Pi(\infty)$ can, using global balance equations (e.g. \cite{stewart2009probability}, be easily calculated (see section Implementation details).

\

We will consider now the case of the step of lenght $\Delta$ with initial probability vector $p(t)$ as in \eqref{silvaepsilon} with the DTMC converging. 

As pointed out in the \cite{Katoen_2006}, in order to rely on the \emph{geometrical convergence} of power iterations for (aperiodic) DTMC the \emph{total variation norm} $l^\infty$, defined as $\Vert \nu \Vert_\infty = max_i\vert \nu_i \vert$ has to be used for the error estimate.

Let us assume that after the $S$ iteration of \eqref{reibmann88pi} ($S < l$ with $l$ for $\epsilon_\Delta$ as in \eqref{reibmann88lk})
\begin{equation}
\label{burak13sse_cond}
\frac{\Vert \Pi(S) - \Pi(\infty) \Vert_\infty}{\Vert \Pi(\infty) \Vert_\infty} < \delta \text{, with }\delta < \varepsilon_T 
\end{equation}
Now instead of further (infinitely) iterating $ \Pi(S)$ in order to calculate $p(t+\Delta)$ as in \eqref{reibmann88}, we will use $\Pi(\infty)$ (instead of $ \Pi(S)$, as proposed in the original algorithm by \cite{muppala1992numerical}) as the $\hat{p}(t+\Delta)$ approximation of such (error free) $p(t+\Delta)$. 

Therefore, for:

\begin{eqnarray}
\label{burak13sserr_1}
p(t+\Delta) - \hat{p}(t+\Delta) & = & \sum_{i=0}^S (\Pi(i)-\Pi(\infty))e^{-\alpha t} \frac{(\alpha t)^{i}}{i!} \\
& + & \sum_{i=S+1}^\infty (\Pi(i)-\Pi(\infty))e^{-\alpha t} \frac{(\alpha t)^{i}}{i!} \nonumber
\end{eqnarray}

as the norm of the first summation is, due to $\Pi(i)$ and $\Pi(\infty)$ being stochastic vectors (and, therefore, $\forall: j \  \vert\Pi_j(i) - \Pi_j(\infty)\vert < 1$) and $S<l$,
strictly upper bounded by:

\begin{equation}
\label{burak13sserr_epsS}
\epsilon_S = Q_\lambda(S) = \sum_{i=0}^S e^{-\alpha t} \frac{(\alpha t)^{i}}{i!} \ \  , \epsilon_S < \frac {\epsilon_\Delta}{2}
\end{equation}
and, as from initial condition:
\[\forall i>S : \Vert \Pi(i) - \Pi(\infty) \Vert_\infty < \delta {\Vert \Pi(\infty) \Vert_\infty}, \text{ and }\sum_{i=S+1}^\infty e^{-\alpha t} \frac{(\alpha t)^{i}}{i!} <1 \]
the resulting error is upper bounded:

\begin{equation}
\label{burak13sserr_1norm}
 \Vert p(t+\Delta) - \hat{p}(t+\Delta) \Vert_\infty < \epsilon_S + \delta {\Vert \Pi(\infty) \Vert_\infty}\ \ , \epsilon_S < \frac{\epsilon_\Delta}{2}
\end{equation}

As steady-state distribution is unique and independent of initial probability distribution (of the step) the above error is absolute and independent of the error of the previous steps.
Therefore:

\begin{equation}
\label{burak13sserr_eps}
\varepsilon_{t+\Delta} = \epsilon_S + \frac{\Vert \Pi(S) - \Pi(\infty) \Vert_\infty}{\Vert \Pi(\infty) \Vert_\infty}
\end{equation}

As the difference (error) between the $\Pi(i)$ and $\Pi(\infty)$ gets smaller, we can expect the angle between both vectors decreasing as well, therefore, the convergence rate will approach asymptotically some value dependent only on subdominant eigenvalue of $P$. If we treat the difference of both vectors as function of current error and some convergence rate function in $i$, we can consequently calculate its value for some $l_S > i$. Although we do not have neither the subdominant eigenvalue nor the exact function of convergence rate $cr(i)$, we can, at any point, easily estimate numerically the current error, the convergence rate and its first derivative, which allows the estimation of $\Vert\Pi(l_S) - \Pi(\infty)\Vert_\infty$ using second order Taylor expansion as:

 \begin{multline}
\label{burak13sserr_Sltaylor}
log\left( \frac{\Vert \Pi(l_S) - \Pi(\infty) \Vert_\infty}{\Vert \Pi(\infty) \Vert_\infty} \right)  =
\\
log\left(\frac{\Vert \Pi(i) - \Pi(\infty) \Vert_\infty}{\Vert \Pi(\infty) \Vert_\infty}\right) + (l_S - i)cr(i) + \frac{(l_S - i)^2}{2!} cr'(i) 
\end{multline}

with the quality of the estimation improving as $lim_{i\rightarrow \infty}(cr'(i)) \rightarrow 0$.

\

Let us assume some $l_S < l$ with $l$ for $\epsilon_\Delta$ as in \eqref{reibmann88lk}, such that $\epsilon_{l_S}$ equal to the cdf of the discrete Poisson distribution function, as in \eqref{burak13sserr_epsS} and insignificant (e.g. $\epsilon_{l_S}=\epsilon_\Delta \times 10^{-3}$). If for some $i < l_S < l$ the relative error of $\Pi(i)$ is greater than the convergence threshold $\delta$ but smaller than some $\delta_T$ ($\delta_T$ relatively small to $\Pi(\infty)$ in order to provide acceptable quality of (\ref{burak13sserr_Sltaylor})) and the relative error of $\Pi(l_S)$ smaller than the convergence threshold $\delta$, then we can stop further iterating $\Pi(i)$ and use $\Pi(\infty)$ as $\hat{p}(t+\Delta)$ with:

\begin{equation}
\label{burak13sserr_eps_taylor}
\varepsilon_{t+\Delta} = \frac{\Vert \Pi(l_S) - \Pi(\infty) \Vert_\infty}{\Vert \Pi(\infty) \Vert_\infty}
\end{equation}

\
However, meeting the convergence can happen earlier, when the error $\varepsilon_t$ from the previous step causes the initial probability vector p(t) being closer to limiting distribution of the current step than the theoretical error-free value. Due to the geometrical convergence of the DTMC this difference is always smaller than:

\begin{multline}
\label{burak13sserr_Sshift}
K_\varepsilon = \frac{log \xi_0 - log(\xi_0 + \varepsilon_t)}{\hat{cr}}, \text{ where: }
\hat{cr}=\frac{log \xi_S - log \xi_0}{S},
\\
\xi_0=\frac{\Vert \Pi(0) - \Pi(\infty) \Vert_\infty}{\Vert \Pi(\infty) \Vert_\infty}, 
\xi_S=\frac{\Vert \Pi(S) - \Pi(\infty) \Vert_\infty}{\Vert \Pi(\infty) \Vert_\infty}, 
\Pi(0)=p(t)
\\
\text{with S replaced by } l_S \text{ if \eqref{burak13sserr_Sltaylor} is used}
\end{multline}
 
Therefore, if the initial (convergence) condition \eqref{burak13sse_cond} has not been met until $S < l-K_\varepsilon$ (or when $l_S+K_\varepsilon > l$ if \eqref{burak13sserr_Sltaylor} is used), $p(t+\Delta)$ will be calculated as in \eqref{reibmann88truncated} and $ \varepsilon_{t+\Delta} = \varepsilon_t + \epsilon_\Delta$. The same applies if the DTMC is not converging.
\\

We can now use the fact that the error bound is, in case the steady state is reached, absolute and independent of the error of the previous steps to set the convergence threshold dependent rather on the  actual total error bound than the error for the single step (as proposed e.g. by \cite{Malhotra_1994}). This is particularly useful in cases of steps with large $\alpha t$ as the relative cost of achieving tighter truncation bonds gets smaller, allowing to trade them for higher convergence thresholds while still within the global error bound for the whole solution.
Assume the system at time $m,0 \leq m <T$. To satisfy $\varepsilon_t < \varepsilon_T$ for each $p(t)$, $t=(m,T]$ we have to:

\begin{equation}
\begin{split}
\label{burak13delta}
\delta_m \leq \varepsilon_T - \varepsilon_m - \sum_m^{T}\epsilon_\Delta
\end{split}
\end{equation}

\subsection{Computational Examples}
To test the implementation the following model has been used: a service system (Call Center) working for time $T$ = 24h and starting empty.  The arrival rate is sinusoidal with two peaks and divided into 288 (5min) periods with constant averaged rates, similar to the example in \citep{Ingolfsson_2010}, with the load varying, in the first example - between 0.65 and 1.05. The service rate  and number of servers  are constant ($\mu(t)=\mu$, $s(t)=s$), the arrival rate varies in time - $\lambda (t) = s \mu (0.85 + 0.2sin(3\pi t / T), 0 \leq t < T$. 

The capacity of the queue is constant and chosen so that for all times the probability $p_n(t)$ of the system being in the state $n$ is less than $1.5 \times 10^{-3}$ for the smallest (30+180) and does not reach $1 \times 10^{-4}$ for all the other systems.

\begin{figure}[h!t!b]
\caption{Computational example 1 - System load}
\label{load}
\centering
\begin{minipage}{0.86\textwidth} 
\includegraphics[width=\linewidth]{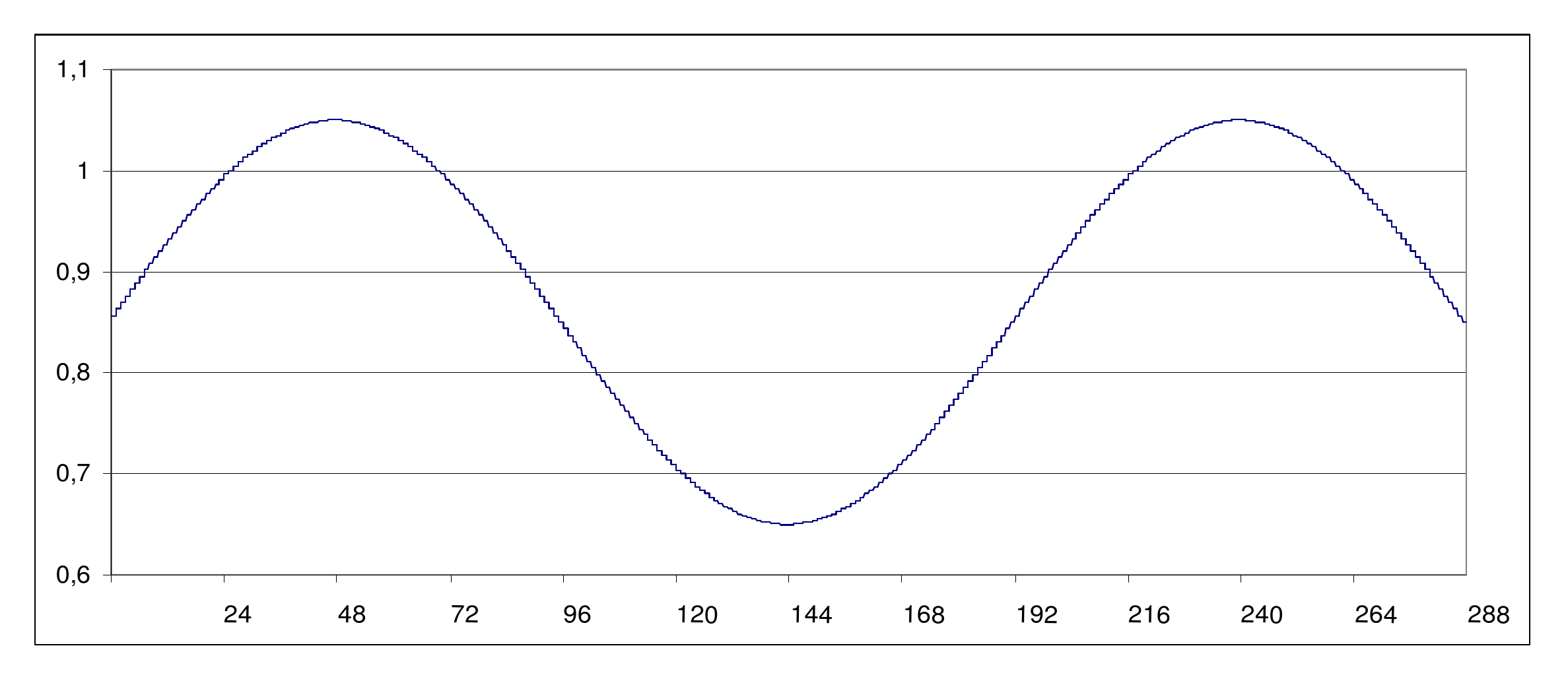}
\end{minipage}
\end{figure}

To evaluate the impact of the proposed steady-state detection algorithm, models of 5 different sizes have been at first calculated  using unmodified uniformization algorithm with an error step $\epsilon=1.5\times 10^{-5}$ corresponding to the total error bound $\varepsilon_T=2.88 \times 10^{-3}$.

All experiments were performed on an 1.7GHz PC under 64bit Linux OS with a processor supporting vector operations in both: avx with 256bit vectors (4 double or 8 float operations simultaneously) and the older sse instruction set with 128bit vector operations (an Intel i5-3317U with cpu throttling disabled via kernel scaling governor), compiled with GNU GCC compiler.

In order to evaluate the impact of our contribution on some common, hardware independent reference (base), the algorithm was firstly implemented using \emph{cblas\_(s/d)gbmv}  (band) matrix vector multiplication from Automatically Tuned Linear Algebra Software (atlas) for GCC compiler as MVM routine.

The working set of the critical part of the implementation requires for every step: access to the initial (e.g. $p(t)$ from previous step) probability vector and creation of new state probability vector which is then used after every MVM to store partial sums and saved "permanently" into RAM after step end.

The calculation of sub-sequential DTMCs requires memory for two probability vectors and 3 rows of the transition matrix. All of them are reused in following steps in order to minimize memory bandwidth. Each of the referenced vectors is of size n+1.

Due to the regular structure of the transition matrix, the second, improved, MVM algorithm calculates consecutive DTMCs creating required operands recursively (on the fly) instead of storing the transition matrix for explicit matrix vector multiplication - consequently reducing working set of MVM loop to only 2 vectors of size n+1.

\begin{table}[h!t!b]
\caption{Comparison base vs. improved MVM algorithm SSE3/AVX double precision.}
\centering
\resizebox{0.75\textwidth}{!}{\begin{minipage}{\textwidth}
\begin{tabular}{l||r|r||r|r||r|r}
$\epsilon_\Delta$=1.5e-5  & \multicolumn{2}{c||}{base ATLAS} & \multicolumn{2}{c||}{improved (sse)} &  \multicolumn{2}{c}{improved (avx)} \\
\hline
System size (s+q) & time & $t/n^2$ & time & $t/n^2$ & time & $t/n^2$ \\
\hline
210........(30+180)&	    67&		1.53&	14&	  	0.32&	6&		0.14	\\
600.......(100+500)&		450&	1.25&	92&		0.26&	39&		0.11	\\
1500.....(300+1200)&		3093&	1.37&	494&	0.21&	224&	0.10	\\
4000....(1000+3000)&		24113&	1.51&	3685&	0.23&	1991&	0.12	\\
9000....(3000+6000)&		155821&	1.92&	22448&	0.28&	12454&	0.15	\\

\end{tabular}
\end{minipage}}

\label{baseSSE3double}
\end{table}

All measurements use standard Unix \emph{time.h/clock()} function - returning CPU time. All times are in milliseconds.

As the precompiled ATLAS library was available only for sse3, all implementations are compared using only this instruction set (compiled with -march=core2 for GCC). The column \emph{improved (avx)} showing the results of the same program compiled using newer instruction set, being clearly an exception from the above statement, only demonstrates that the modified algorithm is vectorizable and scales well with available vector length. 

\

\subsubsection{Steady-state detection}
\

Since the results for the steps with load $\rho \geq 1$ would be identical to those when no steady detection was used, we modified slightly the first example, changing only the arrival rate to $\lambda (t) = s \mu (0.8 + 0.1sin(3\pi t / T), 0 \leq t < T$ with the load $0.7 \leq \rho \leq 0.9$, in order to show the impact of steady state detection on performance and precision in cases when the system can possibly converge.

When applying now \eqref{burak13delta} with e.g. constant $\epsilon_\Delta=1\times 10^{-7}$ and $ \varepsilon_T=0.5\times 10^{-2}$, we can set the steady-state detection threshold $\delta=4.97\times10^{-3}$.

\begin{figure}[h!t!b]
\caption{Number of iterations(mvm) per step, system size 4000 \newline(without convergence error prediction).}
\label{ssd70_90__1000_3000}
\centering
\begin{minipage}{0.86\textwidth} 
\includegraphics[width=\linewidth]{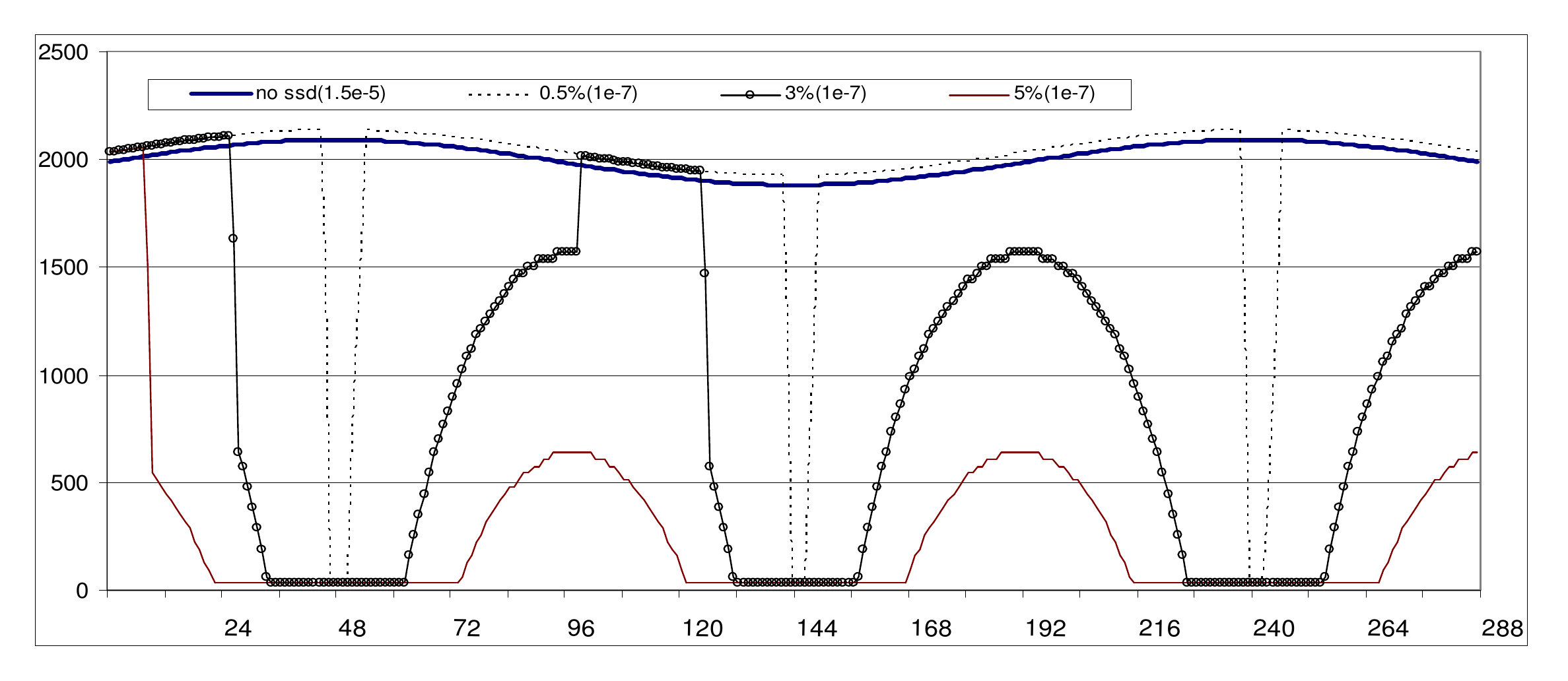}
{\scriptsize load $0.70 \leq \rho \leq 0.90$, s=1000 q=3000\par}
\end{minipage}
\end{figure}

\begin{figure}[h!t!b]
\caption{Number of iterations(mvm) per step, system size 4000 \newline(with convergence error prediction).}
\label{ssd70_90_taylor_1000_3000}
\centering
\begin{minipage}{0.86\textwidth} 
\includegraphics[width=\linewidth]{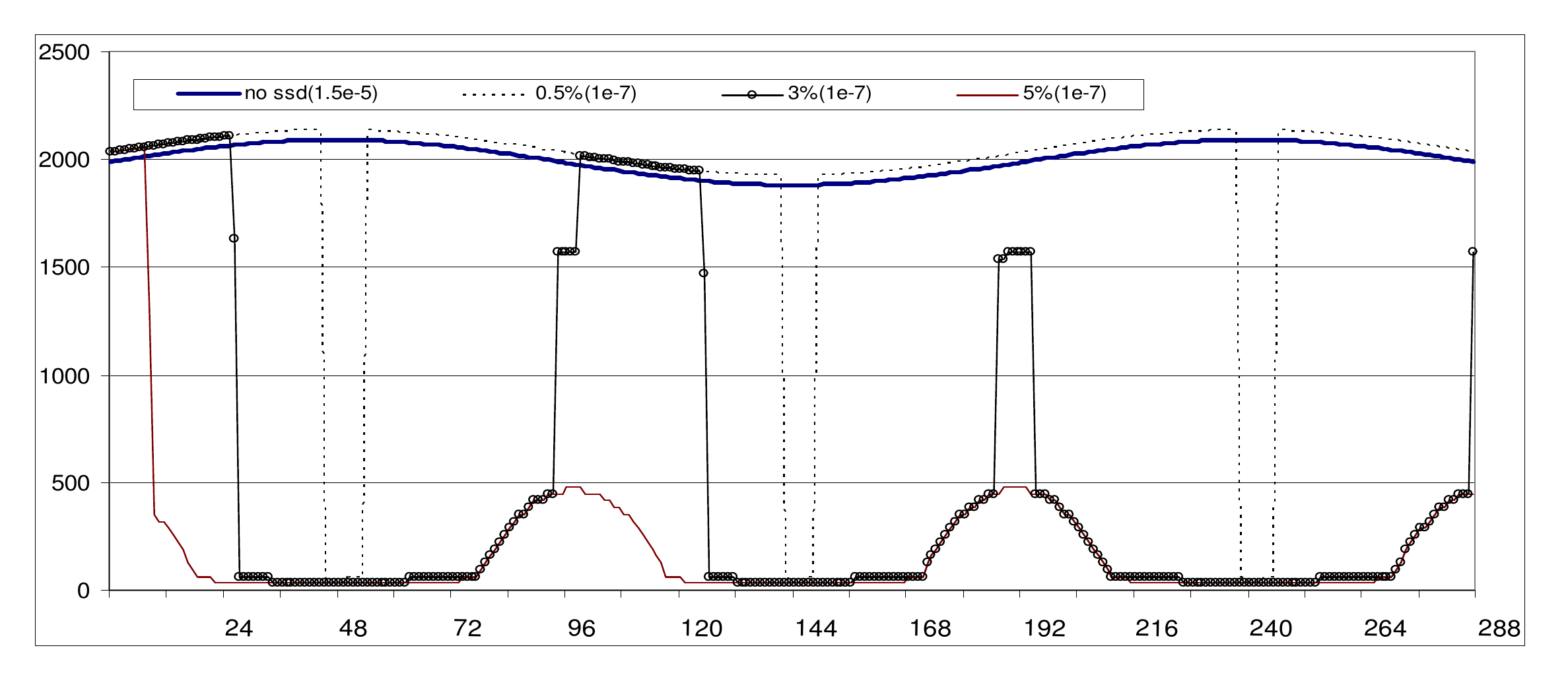}
{\scriptsize load $0.70 \leq \rho \leq 0.90$, s=1000 q=3000,$\delta_T=5.5 \times 10^{-2}$\par}
\end{minipage}
\end{figure}

The impact of reduced computational effort due to steady-state detection for some chosen total error bounds (between $0$ and $5\times10^{-2}$), with corresponding steady-state detection thresholds, is illustrated for the system of size 4000 in Figure \ref{ssd70_90__1000_3000} and with the use of convergence error prediction as in (\ref{burak13sserr_Sltaylor}) in Figure \ref{ssd70_90_taylor_1000_3000} ($\delta_T=5.5 \times 10^{-2}$). 
The detailed results of computation times are in Tables \ref{table_ssd_70_90} and \ref{table_ssd_70_90_taylor}.

\begin{table}[h!t!b]
\caption{Computation times, steady-state detection.}
\resizebox{0.71\textwidth}{!}{\begin{minipage}{\textwidth}
{\begin{tabular}{l||r|r||r|r||r|r||r|r||r|r}
$\epsilon_\Delta$=1e-7 & \multicolumn{2}{c||}{no ssd($\epsilon_\Delta$=1e-5)} & \multicolumn{2}{c||}{$\varepsilon_T=$5e-03} & \multicolumn{2}{c||}{$\varepsilon_T=$1.5e-02} &  \multicolumn{2}{c||}{$\varepsilon_T=$3e-02} & \multicolumn{2}{c}{$\varepsilon_T=$5e-02} \\
\hline
System size & time & $t/n^2$ & time & $t/n^2$ & time & $t/n^2$ & time & $t/n^2$ & time & $t/n^2$ \\
\hline
1500...(300+1200)&		459&	0.20&	404&	0.18&	287&	0.13&	54&		0.024&	33&		0.014\\
4000..(1000+3000)&		3236&	0.21&	3379&	0.21&	2650&	0.17&	1496&	0.094&	424&	0.026\\
9000..(3000+6000)&		20915&	0.26&	21027&	0.26&	19043&	0.20&	16025&	0.198&	9190&	0.113\\
\end{tabular}}
{\scriptsize load $0.7 \leq \rho \leq 0.9$\par}
\end{minipage}}

\label{table_ssd_70_90}
\end{table}

\begin{table}[h!t!b]
\caption{Computation times, steady-state detection with convergence error prediction.}
\resizebox{0.71\textwidth}{!}{\begin{minipage}{\textwidth}
{\begin{tabular}{l||r|r||r|r||r|r||r|r||r|r}
$\epsilon_\Delta$=1e-7,$\delta_T$=5.5e-2 & \multicolumn{2}{c||}{no ssd($\epsilon_\Delta$=1e-5)} & \multicolumn{2}{c||}{$\varepsilon_T=$5e-03} & \multicolumn{2}{c||}{$\varepsilon_T=$1.5e-02} &  \multicolumn{2}{c||}{$\varepsilon_T=$3e-02} & \multicolumn{2}{c}{$\varepsilon_T=$5e-02} \\
\hline
System size & time & $t/n^2$ & time & $t/n^2$ & time & $t/n^2$ & time & $t/n^2$ & time & $t/n^2$ \\
\hline
1500...(300+1200)&		414&	0.19&	402&	0.18&	313&	0.14&	42&		0.018&	33&		0.015\\
4000..(1000+3000)&		3222&	0.20&	3190&	0.20&	2515&	0.16&	876&	0.055&	362&	0.023\\
9000..(3000+6000)&		20492&	0.25&	20542&	0.25&	18299&	0.23&	14449&	0.178&	8015&	0.099\\
\end{tabular}}
{\scriptsize load $0.7 \leq \rho \leq 0.9$\par}
\end{minipage}}
\label{table_ssd_70_90_taylor}
\end{table}

\begin{figure}[h!t!b]
\caption{Number of iterations (mvm) per step, system size 4000 \newline(without convergence error prediction).}
\label{ssd65_105__1000_3000}
\centering
\begin{minipage}{0.86\textwidth} 
\includegraphics[width=\linewidth]{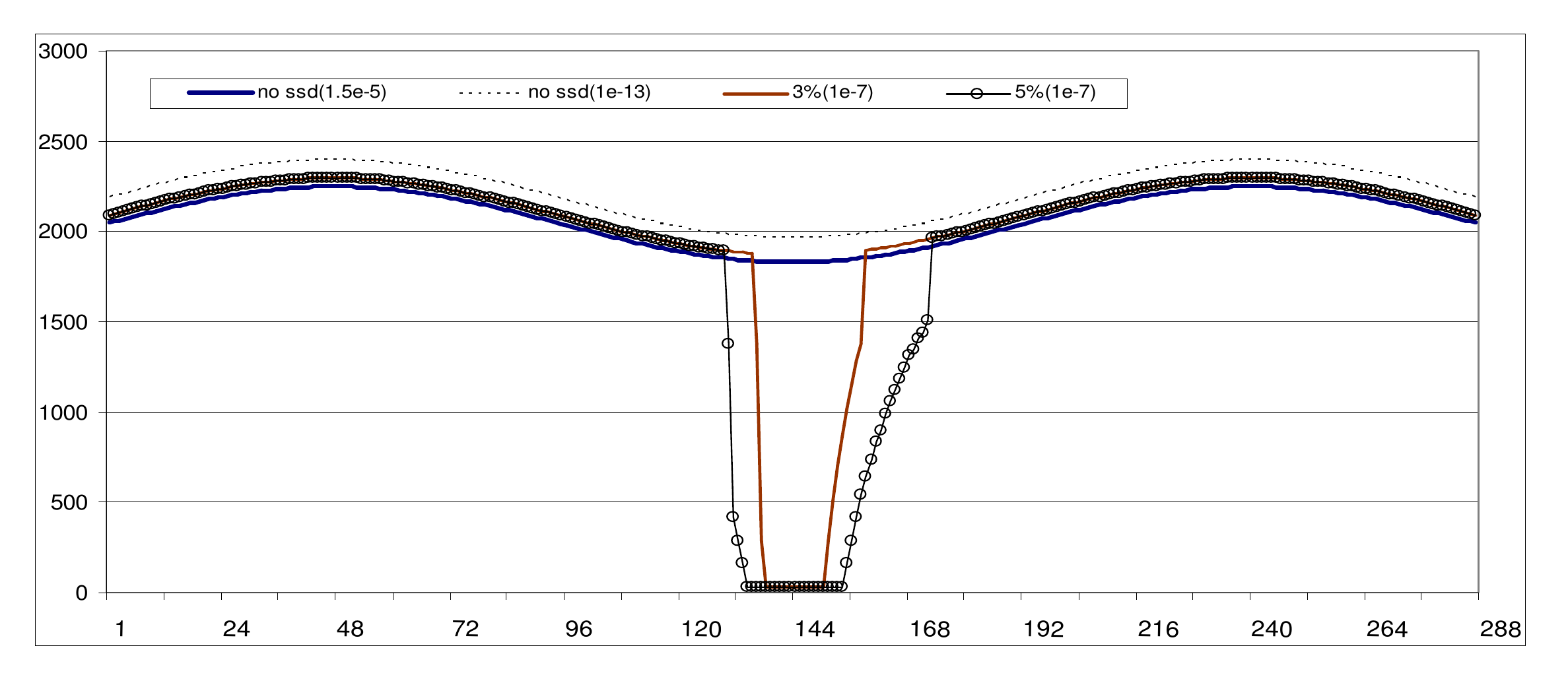}
{\scriptsize load $0.65 \leq \rho \leq 1.05$, s=1000 q=3000\par}
\end{minipage}
\end{figure}

\begin{figure}[h!t!b]
\caption{Number of iterations (mvm) per step, system size 4000 \newline(with convergence error prediction).}
\label{ssd65_105_taylor_1000_3000}
\centering
\begin{minipage}{0.86\textwidth} 
\includegraphics[width=\linewidth]{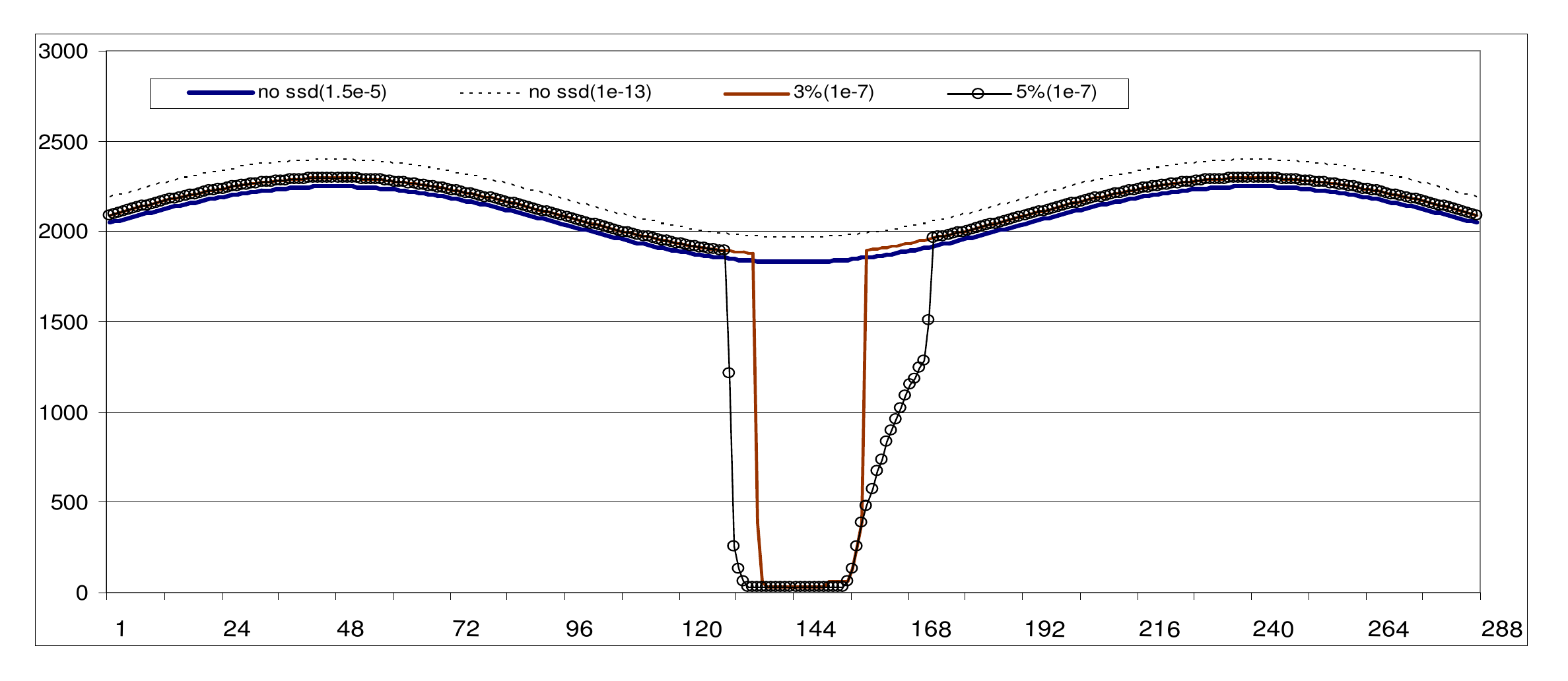}
{\scriptsize load $0.65 \leq \rho \leq 1.05$, s=1000 q=3000\par}
\end{minipage}
\end{figure}

Subsequently, we revisit our first experiment with the load $0.65 \leq \rho \leq 1.05$ (Figure \ref{load}), now using \eqref{burak13delta},  with the same $\varepsilon_T, \epsilon_\Delta$ parameters as in the modified example. The results (with convergence error prediction) are in Table \ref{table_ssd_65_105_taylor} and for the system of size 4000 in Figures \ref{ssd65_105__1000_3000} and \ref{ssd65_105_taylor_1000_3000}.

\begin{table}[h!t!b]
\caption{Computation times, steady-state detection with convergence error prediction.}
\resizebox{0.71\textwidth}{!}{\begin{minipage}{\textwidth}
{\begin{tabular}{p{2.95cm}||r|r||r|r||r|r||r|r||r|r}
$\epsilon_\Delta$=1e-7,$\delta_T$=5.5e-2 & \multicolumn{2}{c||}{no ssd($\epsilon_\Delta$=1e-5)} & \multicolumn{2}{c||}{$\varepsilon_T=$5e-03} & \multicolumn{2}{c||}{$\varepsilon_T=$1.5e-02} &  \multicolumn{2}{c||}{$\varepsilon_T=$3e-02} & \multicolumn{2}{c}{$\varepsilon_T=$5e-02} \\
\hline
System size & time & $t/n^2$ & time & $t/n^2$ & time & $t/n^2$ & time & $t/n^2$ & time & $t/n^2$ \\
\hline
1500...(300+1200)&		429&	0.19&	449&	0.20&	451&	0.20&	396&	0.18&	337&	0.15\\
4000..(1000+3000)&		3364&	0.21&	3530&	0.22&	3419&	0.21&	3313&	0.21&	3207&	0.20\\
9000..(3000+6000)&		21248&	0.26&	21921&	0.27&	21656&	0.27&	21230&	0.26&	20776&	0.26\\
\end{tabular}}
{\scriptsize load $0.65 \leq \rho \leq 1.05$\par}
\end{minipage}}
\label{table_ssd_65_105_taylor}
\end{table}

\begin{figure}[h!t!b]
\caption{Error of the expected system state, system size 4000.}
\label{ssd65_105err__1000_3000}
\centering
\begin{minipage}{0.86\textwidth} 
\includegraphics[width=\linewidth]{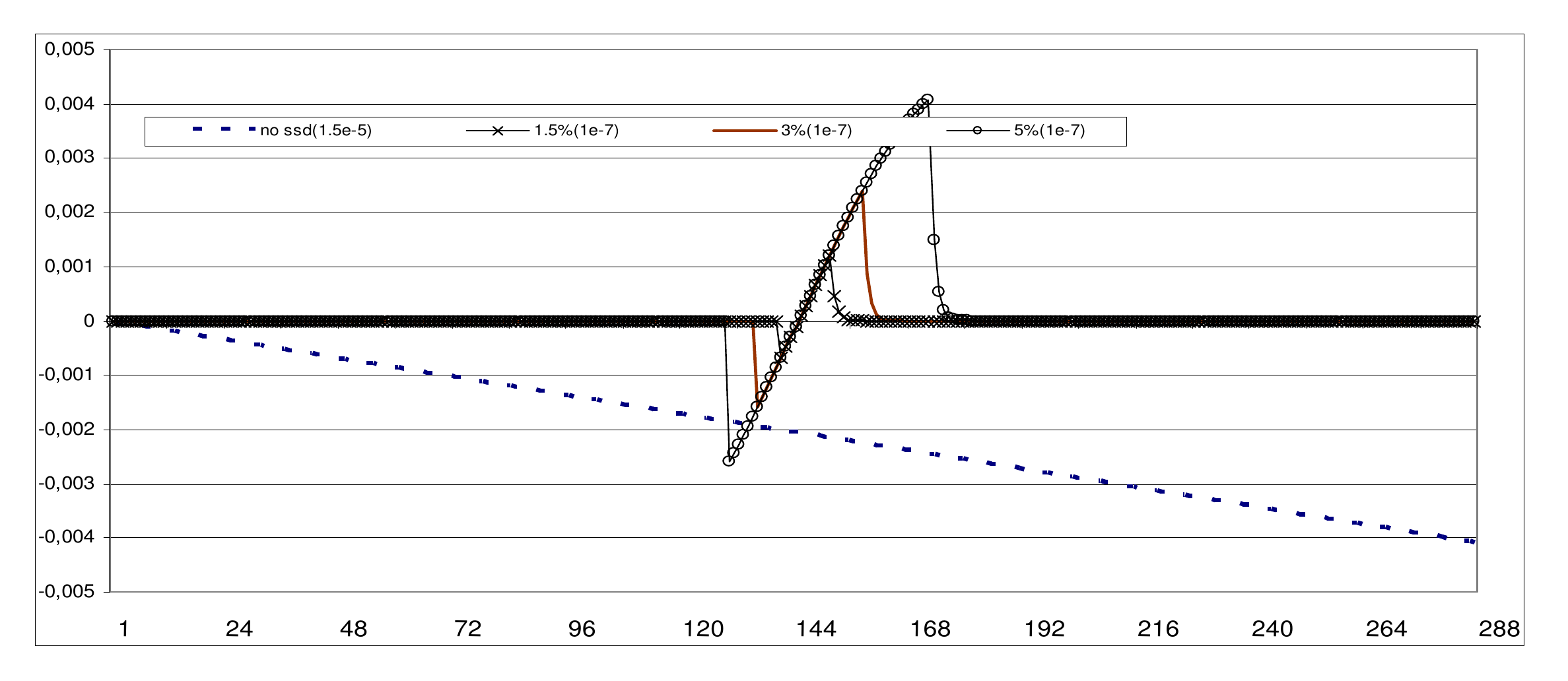}
{\scriptsize load $0.65 \leq \rho \leq 1.05$, s=1000 q=3000\par}
\end{minipage}
\end{figure}

Figure \ref{ssd65_105err__1000_3000} shows the error of the expected state of the system, derived from the calculated probability vector as:
\[ES(t)=\sum_i{i \pi_i(t)},\  p(t)=[\pi_0..\pi_n]\]
The reference for the error estimate has been calculated with $\epsilon_\Delta=1\times10^{-13}$.

\

The computational overhead resulting from steady-state detection (mainly due to the calculation and comparison of vector norms), which has to be done in cases when the system can possibly converge
 ($ \frac{198}{288}$ in the above example) and from tighter step error bounds (traded for the convergence threshold), is offset by savings due to the steady-state detected within the preset error bound, at first for the error bound of $3 \times 10^{-2}$.
  
 However, the savings in the practical Call Center applications should be rather closer to the modified (converging) example, as overloaded time periods will usually constitute much lower percentage of operational times due to the service level requirements.
 The relative increase of computational complexity visible in the $3000+6000$ case in both (original and modified) examples is partially due to filling of the probability vector with very small probability values requiring more floating points operation per MVM. 
 This could be simply ignored for practical reasons, as there is (to the best knowledge of the author) no Call Center on the earth employing more then 1000 agents on a single skill (and even if there were one, then they could probably afford more powerful cpu than the author's $800\$$ notebook). 
 Alternatively, we can consider the probabilities smaller than e.g. $1 \times 10^{-37}$ being insignificant and treat them as if they were equal to $0$, reducing to some extent this effect. The systematic loss of precision due to such truncation would not exceed $\approx 9 \times 10^{-34}$ of the probability mass for the $3000+6000$ case and be, therefore, much smaller than the error due to the truncation of Poisson probability mass (as shown e.g. in the $no ssd$ example on Figure \ref{ssd65_105err__1000_3000}). 
 An additional advantage of this approach is the possibility to store and process the DTMC vectors in single (binary32) precision, which enables up to 2 times more floating point operations in the same time when using sse or up to 4 times more when using avx instruction set. 
 The possible disadvantage is much bigger accumulation of rounding errors due to MVM with much less precision.
 
\begin{table}[h!t!b]
\caption{Computation times, steady-state detection (binary32 on sse).}
\resizebox{0.71\textwidth}{!}{\begin{minipage}{\textwidth}
{\begin{tabular}{l||r|r||r|r||r|r||r|r||r|r}
$\epsilon_\Delta$=1e-7,$\delta_T$=5.5e-2 & \multicolumn{2}{c||}{no ssd($\epsilon_\Delta$=1e-5)} & \multicolumn{2}{c||}{$\varepsilon_T=$5e-03} & \multicolumn{2}{c||}{$\varepsilon_T=$1.5e-02} &  \multicolumn{2}{c||}{$\varepsilon_T=$3e-02} & \multicolumn{2}{c}{$\varepsilon_T=$5e-02} \\
\hline
System size & time & $t/n^2$ & time & $t/n^2$ & time & $t/n^2$ & time & $t/n^2$ & time & $t/n^2$ \\
\hline
1500...(300+1200)&		320&	0.14&	299&	0.13&	194&	0.086&	31&		0.014&	24&		0.011\\
4000..(1000+3000)&		2388&	0.15&	2323&	0.15&	1863&	0.120&	685&	0.043&	232&	0.015\\
9000..(3000+6000)&		14474&	0.18&	14524&	0.18&	12750&	0.160&	10284&	0.126&	5532&	0.068\\
\end{tabular}}
{\scriptsize load $0.7 \leq \rho \leq 0.9$\par}
\end{minipage}}

\label{table_ssd_70_90_sse32}
\end{table}

\begin{table}[h!t!b]
\caption{Computation times, steady-state detection (binary32 on sse).}
\resizebox{0.71\textwidth}{!}{\begin{minipage}{\textwidth}
{\begin{tabular}{l||r|r||r|r||r|r||r|r||r|r}
$\epsilon_\Delta$=1e-7,$\delta_T$=5.5e-2 & \multicolumn{2}{c||}{no ssd($\epsilon_\Delta$=1e-5)} & \multicolumn{2}{c||}{$\varepsilon_T=$5e-03} & \multicolumn{2}{c||}{$\varepsilon_T=$1.5e-02} &  \multicolumn{2}{c||}{$\varepsilon_T=$3e-02} & \multicolumn{2}{c}{$\varepsilon_T=$5e-02} \\
\hline
System size & time & $t/n^2$ & time & $t/n^2$ & time & $t/n^2$ & time & $t/n^2$ & time & $t/n^2$ \\
\hline
1500...(300+1200)&		321&	0.14&	401&	0.18&	336&	0.15&	304&	0.14&	269&	0.12\\
4000..(1000+3000)&		2405&	0.15&	2616&	0.16&	2515&	0.16&	2436&	0.15&	2296&	0.14\\
9000..(3000+6000)&		15319&	0.19&	15738&	0.19&	15413&	0.19&	15296&	0.19&	14920&	0.18\\
\end{tabular}}
{\scriptsize load $0.65 \leq \rho \leq 1.05$\par}
\end{minipage}}

\label{table_ssd_65_105_sse32}
\end{table}

\begin{figure}[h!t!b]
\caption{Error of the expected system state, system size 4000, single (binary32) precision.}
\label{ssd65_105err_float_1000_3000}
\centering
\begin{minipage}{0.86\textwidth} 
\includegraphics[width=\linewidth]{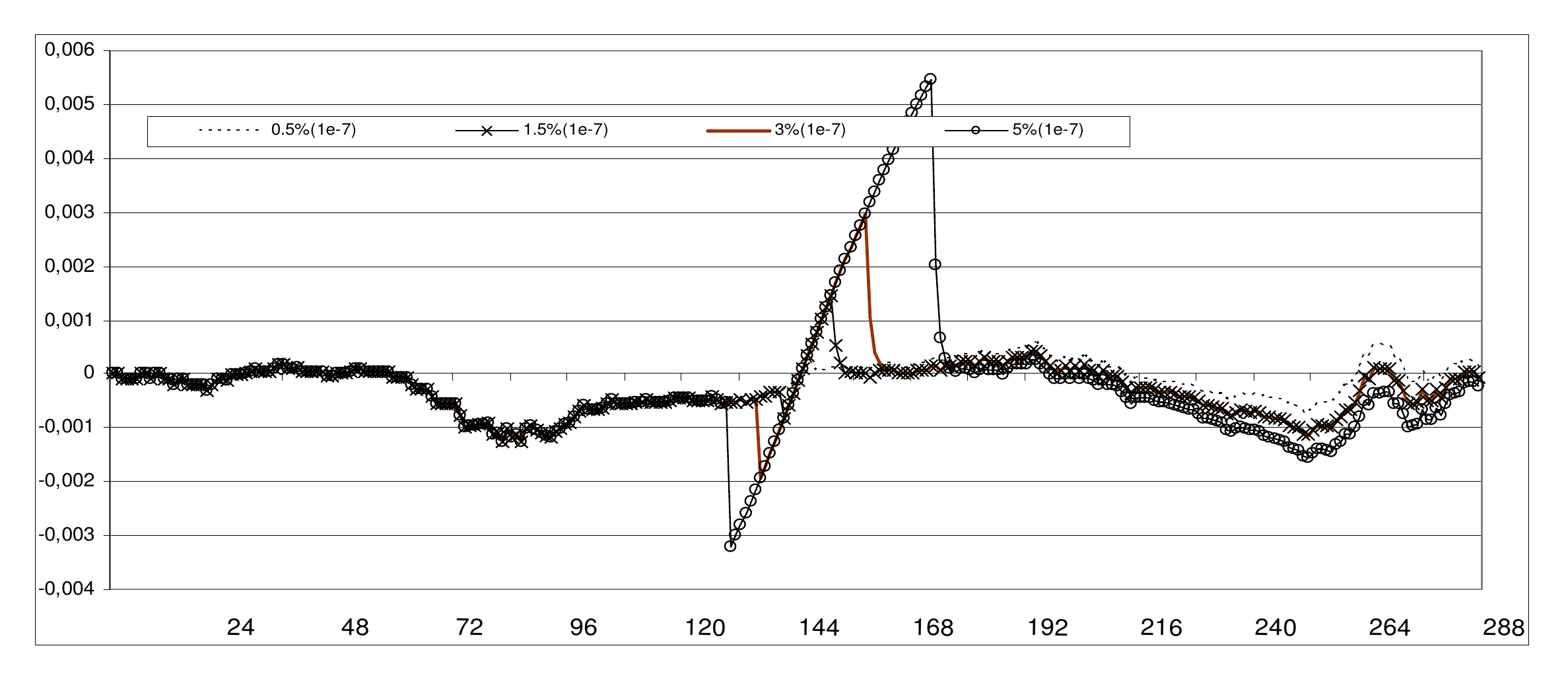}
{\scriptsize load $0.65 \leq \rho \leq 1.05$, s=1000 q=3000\par}
\end{minipage}
\end{figure}

Tables \ref{table_ssd_70_90_sse32} and \ref{table_ssd_65_105_sse32} show execution times of such implementation (for the same data as Tables \ref{table_ssd_70_90}/\ref{table_ssd_70_90_taylor}  and \ref{table_ssd_65_105_taylor}). Figure \ref{ssd65_105err_float_1000_3000} shows its error (similar to Figure \ref{ssd65_105err__1000_3000}). 

\
 
 The results in Tables \ref{table_ssd_70_90_avx32} and \ref{table_ssd_65_105_avx32} show the same computation as the Tables \ref{table_ssd_70_90_sse32} and \ref{table_ssd_65_105_sse32} performed with the newer avx instruction set (on the same hardware). They can not be directly compared with the reference (ATLAS) implementation due to the different instruction set, they show, however, much more efficient utilization of available hardware (using still only 1 of 2 available cpu cores). 
\begin{table}[h!t!b]
\caption{Computation times, with steady-state detection (binary32 on avx).}
\resizebox{0.71\textwidth}{!}{\begin{minipage}{\textwidth}
{\begin{tabular}{l||r|r||r|r||r|r||r|r||r|r}
$\epsilon_\Delta$=1e-7,$\delta_T$=5.5e-2 & \multicolumn{2}{c||}{no ssd($\epsilon_\Delta$=1e-5)} & \multicolumn{2}{c||}{$\varepsilon_T=$5e-03} & \multicolumn{2}{c||}{$\varepsilon_T=$1.5e-02} &  \multicolumn{2}{c||}{$\varepsilon_T=$3e-02} & \multicolumn{2}{c}{$\varepsilon_T=$5e-02} \\
\hline
System size & time & $t/n^2$ & time & $t/n^2$ & time & $t/n^2$ & time & $t/n^2$ &time & $t/n^2$ \\
\hline
1500...(300+1200)&		134&	0.059&	132&	0.059&	86&		0.038&	16&		0.0069&	13&		0.0057\\
4000..(1000+3000)&		1122&	0.070&	1104&	0.069&	894&	0.056&	308&	0.0193&	114&	0.0071\\
9000..(3000+6000)&		6656&	0.082&	6677&	0.082&	5795&	0.071&	4746&	0.0586&	2526&	0.0312\\
\end{tabular}}
{\scriptsize load $0.7 \leq \rho \leq 0.9$\par}
\end{minipage}}

\label{table_ssd_70_90_avx32}
\end{table}
\begin{table}[h!t!b]
\caption{Computation times, with steady-state detection (binary32 on avx).}
\resizebox{0.71\textwidth}{!}{\begin{minipage}{\textwidth}
{\begin{tabular}{l||r|r||r|r||r|r||r|r||r|r}
$\epsilon_\Delta$=1e-7,$\delta_T$=5.5e-2 & \multicolumn{2}{c||}{no ssd($\epsilon_\Delta$=1e-5)} & \multicolumn{2}{c||}{$\varepsilon_T=$5e-03} & \multicolumn{2}{c||}{$\varepsilon_T=$1.5e-02} &  \multicolumn{2}{c||}{$\varepsilon_T=$3e-02} & \multicolumn{2}{c}{$\varepsilon_T=$5e-02} \\
\hline
System size & time & $t/n^2$ & time & $t/n^2$ & time & $t/n^2$ & time & $t/n^2$ &time & $t/n^2$ \\
\hline
1500...(300+1200)&		138&	0.061&	151&	0.067&	146&	0.065&	135&	0.060&	114&	0.051\\
4000..(1000+3000)&		1148&	0.072&	1219&	0.076&	1212&	0.071&	1135&	0.071&	1087&	0.068\\
9000..(3000+6000)&		7312&	0.090&	7585&	0.094&	7492&	0.093&	7345&	0.091&	7271&	0.089\\
\end{tabular}}
{\scriptsize load $0.65 \leq \rho \leq 1.05$\par}
\end{minipage}}

\label{table_ssd_65_105_avx32}
\end{table}

\subsection{Implementation details}

All described algorithms assume that some basic floating point arithmetic is implemented conforming to IEEE 754-2008 standard (e.g. GCC Default IEEE 754 compliance on Intel cpu's supporting either sse or avx instructions). In particular, the numerical accuracy and stability analysis (rounding, underflow and overflow) assumes floating point formats implemented as binary32, binary64 or binary128 (radix 2), correct rounding to the nearest even for basic arithmetic operations and correct conversion from integers. 


\subsubsection{Poisson discrete distribution function}

\

The algorithm presented in Appendix 
computes for given values $\lambda$ ($\alpha t$ in the uniformization algorithm), $\epsilon$ and $\epsilon_{ssd}$: 

the limits $l_{ssd},l,k$, the normalization value $W$ and the weights $w(j), j\in \{l_{ssd} .. k\}$, such that individual Poisson probabilities: $p(i)=w(i)/W , i\in \{l .. k\}$ 

and $w(S)/W = \epsilon_S, S\in \{l_{ssd} .. l-1\}$, where $\epsilon_S$ is equal to the cdf of the discrete Poisson distribution function, as in \eqref{burak13sserr_epsS}. 

If $\epsilon_{ssd}$=$\epsilon$, only the discrete Poisson probabilities are calculated and stored (e.g. in cases of a non converging system), like for the traditional uniformization algorithm.

The algorithm computes the weights recursively starting from mode $m=\lfloor \lambda \rfloor$, with $w(m)$ being the initial weight, similarly to \cite{Fox_1988}. As there is practically no penalty in using double precision numbers in comparison to the single (float) precision for sequential calculations on modern cpu's, the (double precision) binary64 format with $p=53$ bit precision and $w=11$ bit exponent is used. The initial value used for $w(m)$ is 0x1.0p176 ($\approx9.57809713041180536\times10^{52}$).

The calculation of every weight, both left and right to $m$, requires one multiplication and one division.
As we assume basic IEEE 754-2008 conformity, the relative rounding error for every next n'th calculated weight will be upper bounded by  $(1+2^{-53})^{2n-2} -1$.


The rounding error of W is bounded by $(start-end)eps$, where $eps= 2^{-53} \approx 1.11\times10^{-16}$ is the machine epsilon for binary64, but will  probably be much smaller, due to the summation performed in increasing order and the specific properties of the function, as already pointed out in \cite{Fox_1988}, or for much more detailed analysis of the error of sum of exponentially distributed numbers in increased order also  \cite{Robertazzi_1988}. Therefore, the use of more accurate summation algorithms (e.g. Kahan's or Priest's compensated summation) or further increase of precision (e.g. trough using of binary128 type) seems not to be justified.

The number of weights for the calculation of the normalization value $W=\sum_{start}^{end} w()$ is chosen so that the remaining (weighted) cdf is always smaller than the numerical accuracy of $W$ : 
\[Pr(X\leqslant start) + Pr(X\leqslant end) < eps \]
Consequently, $\sum_l^k p() = 1-\epsilon$ and is not always equal to one, as in \cite{Fox_1988}. 
   
\begin{figure}[h!t!b]
\caption{Number of weights to be calculated for given precision as a function of $\lambda$}
\label{B14_weights_L}
\centering
\begin{minipage}{\textwidth} 
\includegraphics[width=0.50\linewidth]{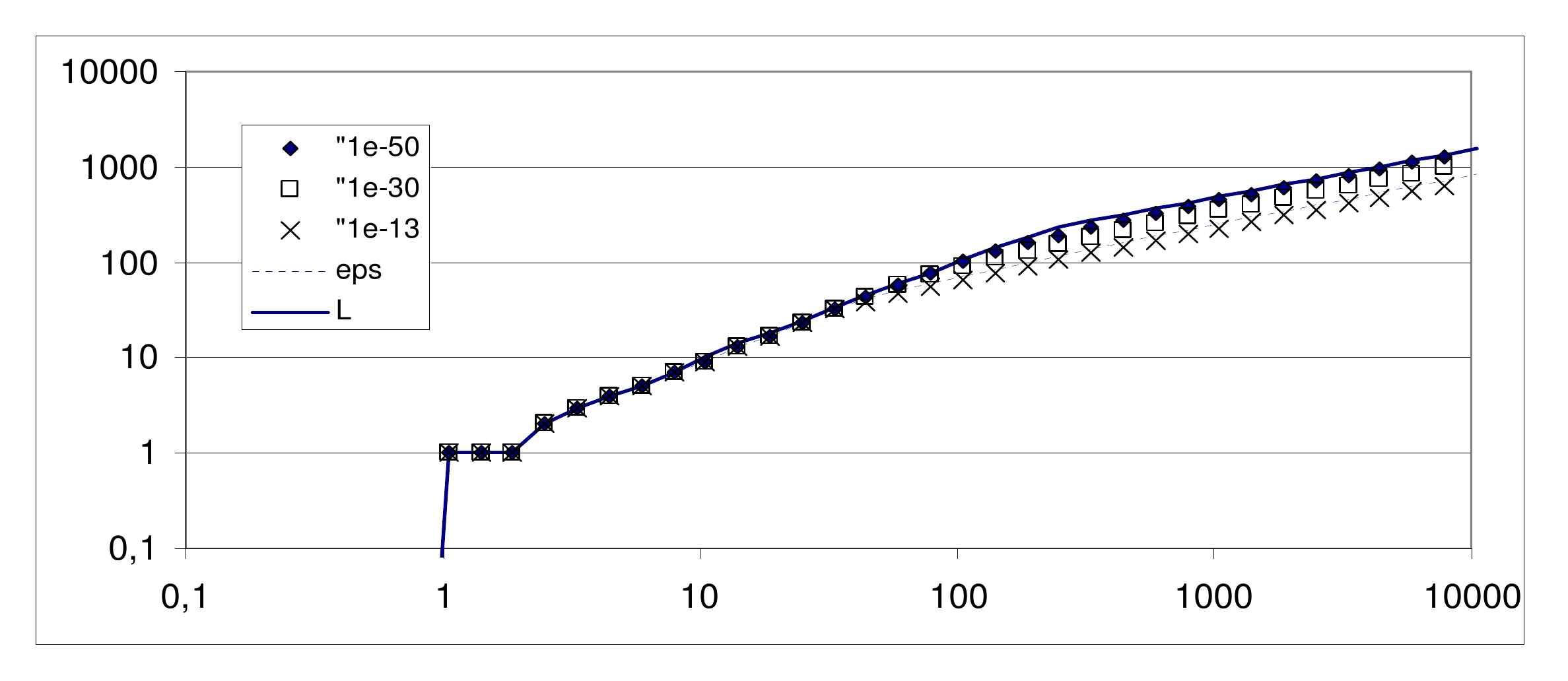}
\includegraphics[width=0.50\linewidth]{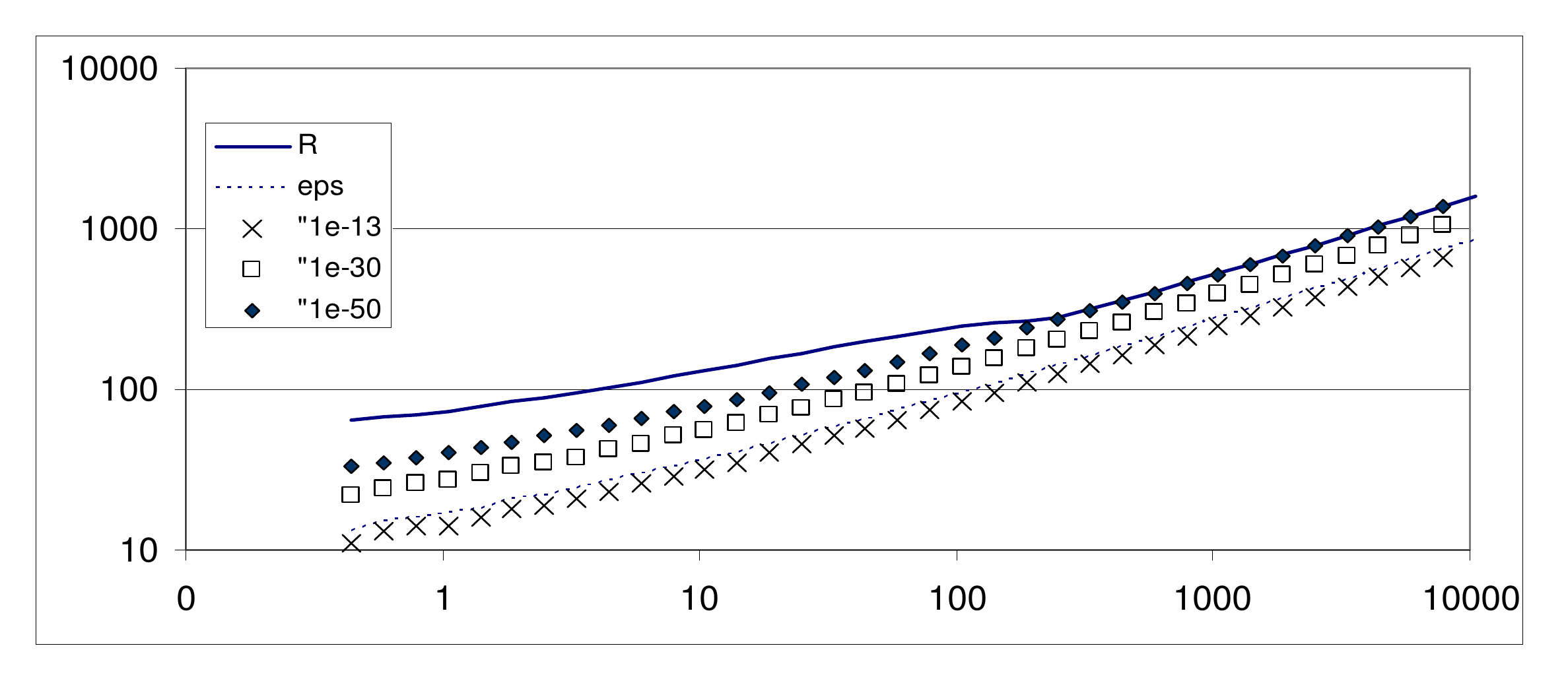}
\end{minipage}
\end{figure}

For bounding the number of calculated weights, the algorithm uses the property of the Poisson distribution having truncation points being asymptotically of $O(\sqrt{\lambda})$ for constant remaining cdf (tail) value and approximates the required number of weights using simple function of $\sqrt{\lambda}$. Figure \ref{B14_weights_L} shows the number of weights necessary to calculate the distribution with truncated probability mass of less than some given value, as a function of $\lambda$, with L and R being the values calculated by the algorithm's approximation. 

The values of weights are (for $0<\lambda<\sqrt{LONG\_MAX} / 2 \approx 1.51\times10^9$ - restricted due to the size of C++ standard \emph{long} format), due to the choice of $w[m]$ and bounding of their number by the approximation, always between $1\times10^{-35}$ and $1\times10^{58}$ and, therefore, comfortably between under- and overflow limits of binary64 format.

The computational complexity is due to the bounding method of $O\sqrt{\lambda}$
 
\subsubsection{M/M/s/n steady state probability vector}

\
The calculation of exact steady state probability vector $\Pi(\infty)$ is done using the well known formula:

\begin{equation}
\label{ErlangC}
\Pi_0^{-1} (\infty )=\sum_{k=0}^{s-1} \frac{(s\rho)^k}{k!} + \sum_{k=s}^n \frac{(s\rho)^k}{s^{k-s}s!}
\ ,\
\Pi_k (\infty )=
\begin{cases}
\displaystyle \Pi_0 (\infty )\frac{(s\rho)^k}{k!}, & 0 < k \leq s\\
 \\
\displaystyle \Pi_0 (\infty )\frac{(s\rho)^k}{s^{k-s}s!}, & s\leq k \leq n
\end{cases}
\end{equation}

The implementation uses its modification similarly to the algorithm calculating Poisson probabilities, with the mode $m = \lfloor s\rho \rfloor$
 and $w[m]=$0x1.0p176. Consequently, the same rounding error approximations for calculating probabilities apart from $m$ apply.
The calculation of weights stops at the limits of the vector or when their values would be no more normalized binary64 numbers using for the remaining weights the value 0x0.0p+0 (to prevent e.g. underflow). The weights are normalized by the total weight to steady state probabilities and only the normalized (either binary32 or binary64) numbers or 0x0p+0 value are stored.

\subsubsection{Uniformization algorithm}

\

Due to the restricted precision of binary32 (float), the probability weighted sum of significant DTMC vectors is done both for binary32 and binary64 implementations in binary64 (double) accuracy. Consequently, the consecutive probability vectors $p(t)$  are always stored and manipulated in binary64 precision (including in particular the casting of precalculated binary32 steady state DTMC vector in cases where steady state was detected). 

\section{Conclusion}
In this paper we showed that the uniformization can be applied in a very effective way to evaluate transient behavior of M/M/n queues. Applied to modeling of the Call Center schedules, it allows calculation of transient system states for systems of any, possible in practical applications, size in a very short time ($\lesssim$1s.), in a numerically stable way, with very high precision, using relatively common and inexpensive CPU. It can, therefore, be used in evaluating possible schedule changes in order to support short-time decisions (e.g. shift times or breaks assignments) in a real-time manner.

Its use can be also extended to schedule planning based on available forecasts, as described in \cite{Ingolfsson_2010}. Particularly useful in this application is the possibility of performance optimization using steady-state detection for systems with load less than 1. As we can start constructing such schedules from an "overstaffed" prototype and then consecutively reduce the applied workforce in order to achieve the desired service level and efficiency, it is then guaranteed that the system will converge for the most of the time during such optimization, reducing the computational time to a fraction, particularly for larger system sizes as shown in the modified (converging) example. It can be even further accelerated by using higher error bounds for first schedule approximations and their step-wise refinement towards a final optimal solution. 


As the proposed error control method is not specific to the queuing systems, it could be also useful  for multi-step solution of other ICTMC's with some homogeneous or almost-homogeneous time steps. Particularly for systems with big state spaces or long step times (large $\alpha t$) , where we can "trade" the then relatively inexpensive higher cdf's truncation precision for higher convergence thresholds, it will result in significant reduction of computational effort without compromising strict error bounds for the whole (multi-step) solution.

The presented method can be extended in several directions. The first one could be, in regard to Call Center modeling, to consider abandonment due to impatient customers or more generally consider cases where we can easily calculate the precise steady-state of the system  in advance(e.g using flow equations). Another would be to apply it in the similar way to other inhomogeneous CTMCs in cases, where steady state for a step could be efficiently approximated using faster converging iterative methods (e.g. Lanczos) parallel to the DTMC evolution.

Finally, as the presented algorithm is quite thrifty in memory bandwidth use and at the same time its split for parallel calculation of (parts of) consecutive DTMC probability vectors would need relatively little synchronization due to the on the fly generation of coefficients, it could profit from multi-threaded execution on separate CPU cores in an almost linear way (e.g. execution on a 8-core CPU would require almost 8 times less time). This could be relatively easily implemented  using, for example, the standard OpenMP extension.

\nocite{fog2006optimizing}



\newpage
\section*{APPENDIX: Class B14 - Computing Poisson Probabilities}
 \lstset{language=C++,
            basicstyle=\scriptsize,
          breaklines=true
          }
\begin{lstlisting}

#ifndef Poisson_Burak2014
#define Poisson_Burak2014

#include<cmath>
#include<cstdlib>

class B14 {
   private:     long wsize,start;
                long left_l,right_k;
                double *weights;
                double total_weight;
   public:
//constructor - creates weights    
  B14(const double Lambda, double Epsilon, double Epsilon_ssd); 

  long LS(){return start;}   //left truncation point for ssd epsilons
  long L() {return left_l;}  //left truncation point
  long R() {return right_k;} //rigt truncation point

//weights value by the n-th poisson probability
  double p(const double value, const long n)      
   {if(value > total_weight)
    return((n>=left_l)&&(n<(wsize+start)) ? 
  			(value/total_weight)*weights[n-start] : 0.0);
   return((n>=left_l)&&(n<(wsize+start)) ? 
   		(value*weights[n-start])/total_weight : 0.0);}

//returns S-th cdf    
  double essd(const long S)       
   {return ((S>=start) && (S<left_l) ? 
   		weights[S-start]/total_weight : 0.0);}

//returns n-th weight, use together with W()
  double weight(const long n)  
   {return ((n>=start) && (n<wsize+start) ? 
   		weights[n-start] : 0.0);}

//for weighted sums of small values e.g. probabilities: 
//					sum_i( weight(i) * p_i ) / W()
  double W(){return total_weight;}
  
  ~B14(){free(weights);}
  
}; //end of B14 class


B14::B14(const double Lambda, double Epsilon, double Epsilon_ssd)
    {if(!(Lambda > 0.0)||!(Epsilon<1.0d))
      {left_l=right_k=0;weights=&total_weight;
       total_weight=nan("");return;}
     if(Epsilon<1e-50)Epsilon=1e-50; 
     if(Epsilon_ssd<1e-50)Epsilon_ssd=1e-50;
     if(!(Epsilon_ssd<Epsilon))Epsilon_ssd=Epsilon;  //LS=L essd=0.0

     long m=(long)floor(Lambda);
     
//number of temporary weights stored
     long mw=30; long ma=44; long ms=21;
     long tsize=(long)(sqrt(Lambda) * mw)  + ma;
     long tstart=m+ms-tsize/2;if(tstart<0)tstart=0;
     
//if short of stack and big Lambda: alloca -> malloc+free
     double *tweights=(double*)alloca(tsize*sizeof(double)); 

     long j=m-tstart;
     tweights[j]=0x1.0p176; //weight[m]

     for(j=m-tstart;j>0;j--)
     	tweights[j-1]=(tweights[j]*(j+tstart))/Lambda;
     for(j=m-tstart+1;j<tsize;j++)
     	tweights[j]=(Lambda*tweights[j-1])/(j+tstart);

//compute total_weight
     total_weight=0.0;
     for(j=0;j<(m-tstart);j++)total_weight+=tweights[j];
     double suml=0.0;
     for(j=(tsize-1);j>=(m-tstart);j--)suml+=tweights[j];
     total_weight+=suml;

//calculate truncation points
     double ogon=(Epsilon_ssd*total_weight)/2.0;
     long i=0;
     double cdf=tweights[i];
     while(cdf<ogon) cdf+=tweights[++i];
     start=i+tstart;
     double cdf_start=cdf;

     ogon=(Epsilon*total_weight)/2.0;
     while(cdf<ogon) cdf+=tweights[++i];
     left_l=i+tstart;
       //if(i==0)ogon*=2; //for historical compatibility
     i=tsize-1;
     cdf=tweights[i];
     while(cdf<ogon) cdf+=tweights[--i];
     right_k=i+tstart;

//only weights/cdfs between LS and R are stored
        wsize=right_k-start+1;
        weights = (double *)malloc(wsize * sizeof(double));
        weights[0]=cdf_start;

     for(j=start+1;j<left_l;j++)
       weights[j-start]=weights[j-start-1]+tweights[j-tstart];
     for(j=left_l;j<=right_k;j++)
       weights[j-start]=tweights[j-tstart];
}//end of B14 constructor

#endif
\end{lstlisting}

\newpage

\section*{References}

\end{document}